# ∑3 (111) Grain Boundaries Accelerate Hydrogen Insertion into Palladium Nanostructures

K. A. U. Madhushani,[1†] Hyoju Park,[2†] Hua Zhou,[3] Diptangshu Datta Mal,[1] Qin Pang,[2]

Dongsheng Li,[2*] Peter V. Sushko,[2*] Long Luo[1*]

1. Department of Chemistry, University of Utah, Salt Lake City, Utah 84112, United States
2. Physical & Computational Sciences Directorate, Pacific Northwest National Laboratory, Richland, Washington 99354, United States
3. X-Ray Science Division, Argonne National Laboratory, Lemont, Illinois 60439, United States

**ABSTRACT:** Grain boundaries (GBs) are frequently implicated as key defect structures facilitating metal hydride formation, yet their specific role remains poorly understood due to their structural complexity. Here, we investigate hydrogen insertion in Pd nanostructures enriched with well-defined ∑3(111) GBs ($Pd_{GB}$) synthesized via electrolysis-driven nanoparticle assembly. In situ synchrotron X-ray diffraction reveals that $Pd_{GB}$ exhibits dramatically accelerated hydriding and dehydriding kinetics compared to ligand-free and ligand-capped Pd nanoparticles with similar crystallite sizes. Strain mapping using environmental transmission electron microscopy shows that strain is highly localized at GBs and intensifies upon hydrogen exposure, indicating preferential hydrogen insertion along GB sites. Density functional theory calculations support these findings, showing that hydrogen insertion near ∑3(111) GBs is energetically more favorable and that tensile strain lowers insertion barriers. These results provide atomic-level insights into the role of GBs in hydride formation and suggest new design strategies for GB-engineered Pd-based functional materials.

## Introduction

Metal hydride formation is a critical process because it underpins a wide range of energy technologies and phenomena, including hydrogen storage,[1-3] metal-hydride electrochemical energy storage,[4-7] electrochemical energy conversion (e.g., using fuel cells),[3] chemical transformations including hydrogenation and dehydrogenation,[8,9] hydrogen gas sensing,[10-17] hydrogen purification,[18,19] and degradation of energy systems via hydrogen embrittlement.[20-24] Understanding the interaction of hydrogen with metals is also important in the context of hydrogen trapping in fusion materials and controlling plasma processing in microelectronics devices. [25-28]

Pd is widely studied for its metal–hydrogen interactions due to its high hydrogen absorption capacity and fast kinetics at ambient conditions. Early studies focused on macroscopic Pd materials, such as single crystals and polycrystalline films.[29-34] However, due to the high cost of Pd and, accordingly, the need to utilize Pd more effectively, research has shifted toward understanding hydride formation at the nanoscale, particularly the effects of the size and shape of nanoparticles (NPs). Yamauchi et al. showed that smaller Pd NPs have stronger Pd–H interactions compared with macroscopic Pd.[35] Ingham et al. confirmed similar size effects using in situ synchrotron X-ray diffraction (XRD),[36] while Langhammer et al. demonstrated diffusion-limited hydriding kinetics and surface tension-controlled hydrogen desorption for Pd NPs using nanoplasmonic sensing.[37] Berlinguette et al. found that lattice-strained vertices in Pd nanocrystals accelerate hydrogen uptake.[38] Meanwhile, Zlotea et al. observed a size-induced phase transition from fcc to icosahedral structures in ~2.5 nm Pd NPs using XRD and Extended X-ray absorption fine structure (EXAFS).[39] Lamberti et al. reported core–shell behavior, where a crystalline core undergoes sharp α–β transitions while an amorphous shell absorbs hydrogen gradually.[40] Dionne et al. used in situ STEM to track β-phase nucleation in Pd nanocubes, starting at corners and propagating along (100) planes, with size-dependent equilibrium pressures.[41,42] Monte Carlo simulations by Ruda et al. showed that different NP shapes lead to multi-plateau behavior in pressure-composition isotherms, unlike bulk Pd.[43] Recently, interest has expanded to bimetallic systems. Kitagawa et al. used solid-state $^2H$ NMR to show hydrogen accumulation at Pd/Pt core–shell interfaces, highlighting the role of interfacial sites in hydride stabilization.[44] Ogura et al. found that submonolayer Au alloying on Pd(110) enhances hydrogen uptake by lowering the insertion barrier, attributed to electronic effects observed via ARPES.[45]

While the effects of particle size, shape, and composition on Pd hydride thermodynamics and kinetics are relatively well understood, the role of structural defects, particularly grain boundaries (GBs), remains less clear. This gap is especially notable given that GBs are often considered key sites for hydrogen absorption, diffusion, and phase transitions.[46-48]

Studying hydride behavior at GBs is challenging due to their structural complexity. Each GB is defined by five degrees of freedom: three for grain misorientation and two for boundary plane orientation. This complexity makes it difficult to isolate specific GB types and establish clear structure–property relationships. Only a few studies have explored this topic. For example, atomistic simulations suggest GBs in Ni promote hydride formation due to favorable hydrogen binding sites.[23] In Pd, modeling predicts enhanced hydrogen storage in GB-rich nanostructures,[49] and experimental methods like cryo atom probe tomography[20] and plasmonic imaging[50, 51] have started to link GB structures with hydride kinetics.

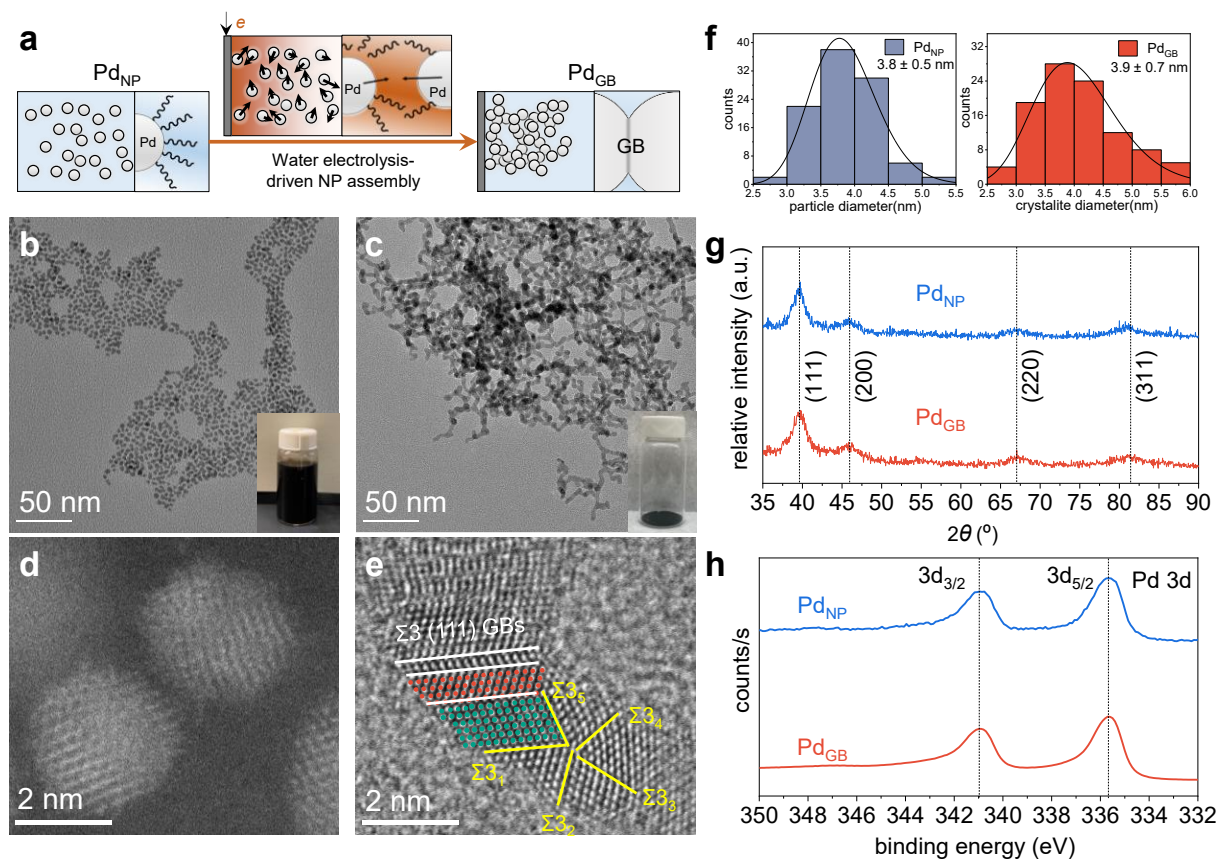

**Figure 1. Synthesis and characterization of Pd nanoparticles (Pd$_{NP}$) and their grain boundary (GB)-rich assembly (Pd$_{GB}$).** (a) Schematic of the water electrolysis-driven assembly of Pd$_{NP}$ into Pd$_{GB}$. (b, d) TEM images of Pd$_{NP}$. (c, e) TEM images of Pd$_{GB}$, showing the prevalence of Σ3(111) GBs and a five-fold Σ3 GBs (Σ3$_1$ to Σ3$_5$). Red and green dots represent individual atoms in two domains separated by a Σ3(111) GB. (f) Crystallite size distributions, (g) XRD patterns, and (h) XPS spectra of Pd$_{NP}$ and Pd$_{GB}$.

To address the lack of the structure-property relationships for metal hydride formation at GBs, we synthesized samples in the form of interconnected quasi-1D Pd chains with well-defined Σ3(111) GBs, which not only allow for a detailed characterization of the GB's structural changes upon hydrogenation but also enables novel responses to H uptake, using in situ XRD and in situ TEM, and complementary computational modeling. Experimentally, we observe a rapid phase transition from Pd to PdH$_x$ and back, with strain mapping revealing localized strain at these GBs. In situ HR-TEM results suggest that hydrogen insertion prefers the Σ3(111) GBs over non-GB sites, confirming their role in facilitating hydride formation. DFT calculations suggest that hydrogen preferentially accumulates near the concave sites in the vicinity of GBs and further show that tensile strain at GBs lowers the hydrogen insertion barrier.

### Results and Discussion

The Pd nanostructure enriched with Σ3(111) GBs (Pd$_{GB}$) was synthesized via a water electrolysis-driven Pd NP (Pd$_{NP}$) assembly method previously developed by our group (illustrated in **Figure 1a**).[52] This strategy utilizes electrogenerated H$_2$ and the localized high pH near the cathode to remove citrate ligands from Pd NPs, destabilizing them and promoting random collisions in solution.[52] These collisions result in oriented attachment predominantly at the (111) facets,[53] leading to the formation of Σ3(111) GBs.[54]

**Figures 1b–e** show TEM images and optical photographs of Pd$_{NP}$ and their GB-rich assemblies (Pd$_{GB}$). High-resolution TEM images in **Figures 1e** and **S1-3** reveal the presence of parallel Σ3(111) GBs and several five-fold Σ3 GBs (Σ3$_1$ to Σ3$_5$) where five grains are arranged in a cyclic structure around a common axis (See **supporting information (SI)** for details on GB identification). The five-fold twins are believed to form through repeated oriented attachment of NPs. This configuration promotes the creation of additional Σ3(111) GBs through high-energy GB decomposition or partial dislocation slipping.[54] Earlier studies indicated that approximately 15% of oriented attachment events in Pd NPs would result in the formation of five-fold twins.[54]

TEM-based size analysis reveals that Pd$_{NP}$ and Pd$_{GB}$ have similar crystallite size distributions: 3.8 ± 0.5 nm for Pd$_{NP}$ and 3.9 ± 0.7 nm for Pd$_{GB}$ (**Figure 1f** and **S4**). This finding is further supported by comparable XRD peak widths for both samples (**Figures 1g** and **S5**). Using the

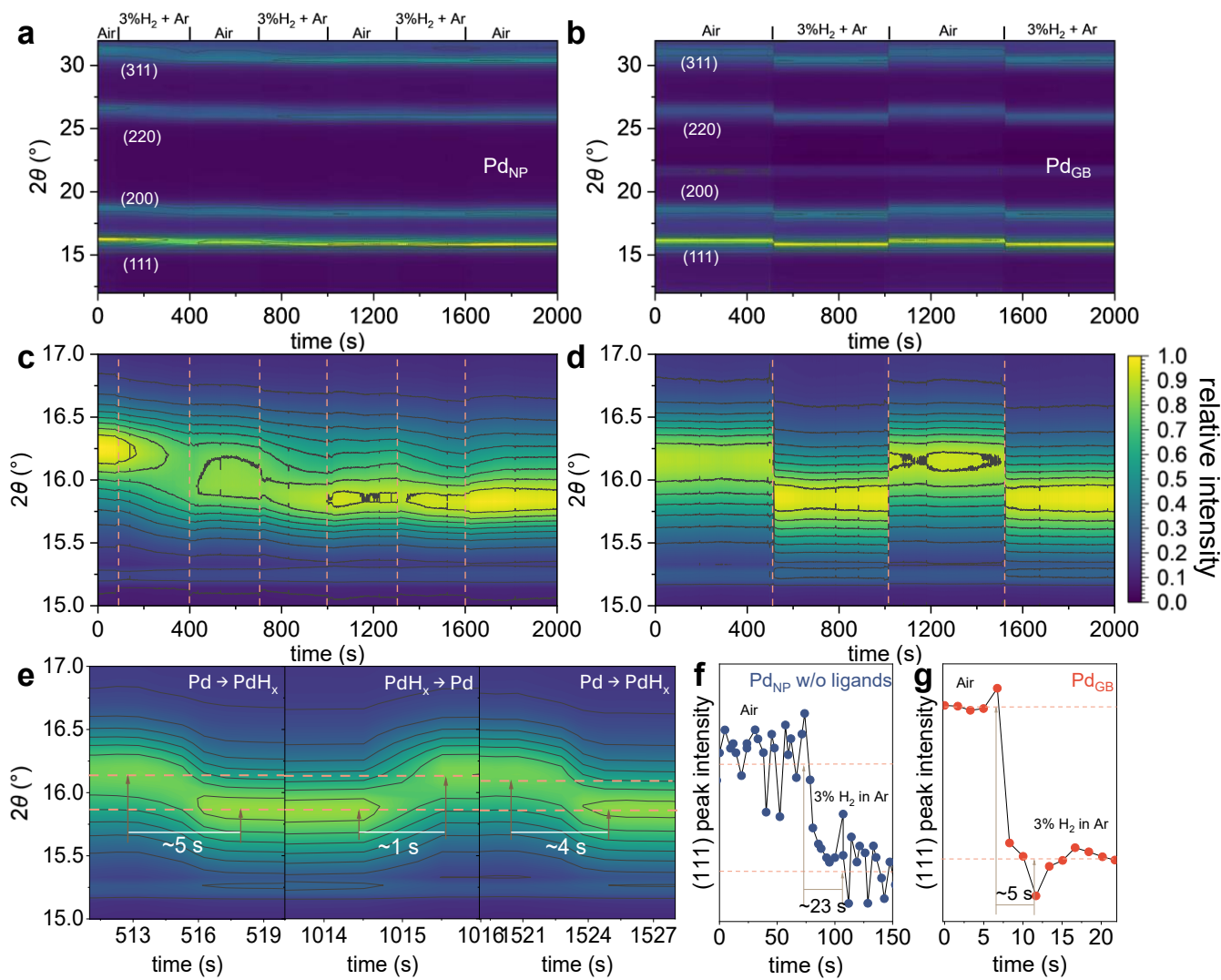

**Figure 2. In situ synchrotron XRD analysis of hydrogen-induced phase transitions.** (a, b) Time-resolved XRD patterns of Pd$_{NP}$ and Pd$_{GB}$ under alternating flows of air and 3% H$_2$ in Ar. (c, d) Enlarged views of the Pd(111) diffraction peak for Pd$_{NP}$ and Pd$_{GB}$. The dashed lines indicate the switching points between air and H$_2$/Ar gas flows. (e) Time-resolved phase transition dynamics between Pd and PdH$_x$ in the presence and absence of 3% H$_2$ in Ar for Pd$_{GB}$. (f, g) Comparison of the Pd(111) peak intensity changes between ligand-free Pd$_{NP}$ and Pd$_{GB}$ upon exposure to 3% H$_2$ in Ar.

Scherrer equation, the average crystallite sizes were calculated to be 3.6 nm for Pd$_{NP}$ and 3.8 nm for Pd$_{GB}$ from the (220) peak of their high-resolution XRD pattern (see **SI** for details). The similarity in crystallite size is crucial for this study, as the rate of Pd hydride formation is known to depend strongly on particle size.[35-37] By ensuring comparable crystallite sizes, we can isolate and evaluate the influence of other structural features, such as GBs, on hydride formation behavior. The XPS spectra of Pd 3d in **Figure 1h** also confirm that Pd in both samples exists in the metallic state (Pd$^0$).

Next, we performed *in situ* synchrotron XRD analysis to examine the transition of Pd$_{NP}$ and Pd$_{GB}$ to their respective hydride phases. **Figures 2a** and **2c** display the time-resolved XRD patterns of Pd$_{NP}$ under alternating flows of air and 3% H$_2$ in Ar. The Pd (111) peak, initially located at ~16.2°, gradually shifted to ~16.0° after 5 minutes of H$_2$ exposure, indicating the slow formation of PdH$_x$. Upon switching back to air, the peak location showed minimal reversal, suggesting sluggish hydrogen desorption, likely due to citrate surface ligands on Pd$_{NP}$ that impede hydrogen atom recombination and, consequently, hydrogen desorption. This irreversible absorption/desorption behavior resulted in a progressive, one-directional negative shift of the Pd(111) peak, stabilizing at ~15.8° after several H$_2$ on/off cycles. The overall lattice expansion of ~2.3% in 3% H$_2$ is close to the previously reported value of ~2.5% under similar conditions,[40] which corresponds to a mixed α and β phases of PdH$_x$. In stark contrast, Pd$_{GB}$ exhibited a rapid and reversible XRD peak shift in response to gas composition changes (**Figures 2b** and **2d**). As shown in **Figure 2e**, the Pd-to-PdH$_x$ transition, corresponding to the Pd(111) peak shift from 16.1° to 15.8°, was completed within ~4–5 seconds upon exposure to 3% H$_2$, and the reverse transition back to metallic Pd occurred within just 1 second after switching to air.

To isolate the effect of surface ligands on hydrogen absorption and desorption, we prepared ligand-free Pd$_{NP}$ by treating the original citrate-capped Pd$_{NP}$ with UV-ozone.

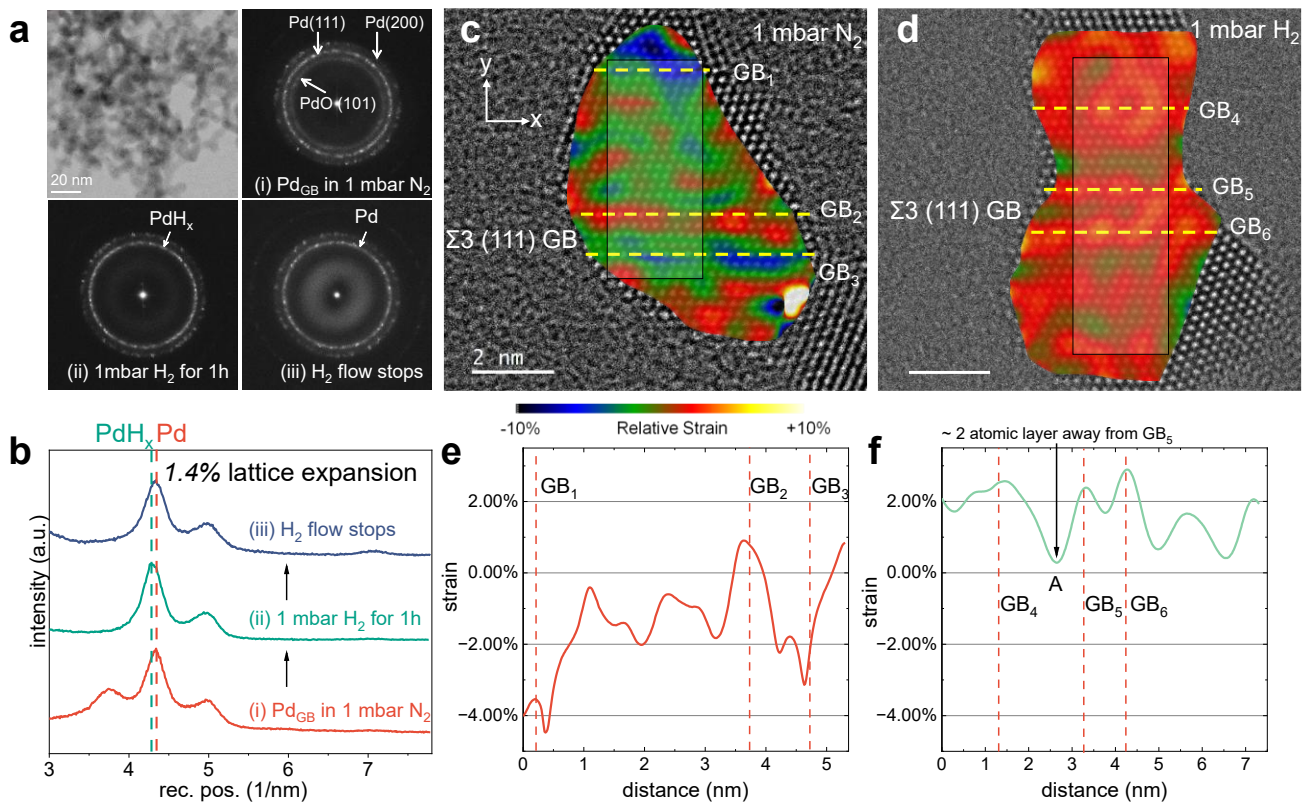

**Figure 3. In situ environmental TEM analysis of hydrogen-induced phase transitions.** (a) Fast Fourier Transform (FFT) patterns of Pd$_{GB}$ under the following sequential conditions: (i) 1 mbar N$_2$, (ii) 1 mbar H$_2$, and (iii) after stopping H$_2$ flow. (b) Integrated intensity profiles of the FFT patterns shown in panel (a), plotted as a function of reciprocal space position. (c, d) HRTEM images and corresponding y-direction strain maps of Pd$_{GB}$ under (c) 1 mbar N$_2$ and (d) 1 mbar H$_2$. (e, f) Average strain profiles as a function of position for the boxed regions in panels (c) and (d). The minimum strain point A is ∼ 2 atomic layers away from GB$_5$.

This treatment selectively removes organic surface impurities without compromising the structural integrity of the Pd nanocrystals,[53, 55] as confirmed by XRD, XPS, and TEM (**Figures S6-S8**). Following ligand removal, the hydrogen-induced phase transition became fully reversible (**Figure S9**), confirming that citrate ligands indeed inhibit hydrogen desorption. Moreover, the phase transition is significantly accelerated relative to the original, ligated Pd$_{NP}$. **Figure 2f** shows the change in Pd (111) XRD peak intensity for the ligand-free Pd$_{NP}$ upon exposure to 3% H$_2$ in Ar. The phase transition was completed in approximately 23 seconds—substantially faster than the ligated sample, but still much slower than the ∼5-second transition observed for Pd$_{GB}$ (**Figure 2g**). These results support that Σ3(111) GBs play a key role in facilitating rapid hydrogen insertion into Pd.

To elucidate the relationship between Pd structure, Pd Σ3(111) GBs content, and hydrogen insertion, we performed in situ TEM experiments on the Pd$_{GB}$ sample. **Figure 3a** presents three sequential FFT patterns collected under 1 mbar N$_2$, after 1 h exposure to 1 mbar H$_2$, and after H$_2$ removal. The initial FFT pattern displays three diffraction rings, corresponding to PdO(101), Pd(111), and Pd(200) (from inner to outer). The presence of PdO likely originates from the gradual surface oxidation of Pd$_{GB}$ upon prolonged air exposure before the in-situ TEM study. Upon H$_2$ exposure, the innermost ring, attributed to cubic PdO(101), disappears. The integrated FFT intensity profile in **Figure 3b** shows the corresponding disappearance of the ∼3.745 nm$^{-1}$ peak (d-spacing = 0.267 nm, tetragonal PdO(101)).[56] Simultaneously, the primary Pd(111) peak at 4.340 nm$^{-1}$ (d-spacing = 0.230 nm) shifts to 4.283 nm$^{-1}$ (d-spacing = 0.233 nm), indicating lattice expansion due to PdH$_x$ formation. After stopping the H$_2$ flow, the Pd(111) peak shifts back to its original position, confirming the reversible phase transition. The observed shift corresponds to an approximate 1.4% lattice expansion, which is higher than the previously reported value of ∼0.2% for NPs with similar size under similar H$_2$ pressure.[40] This discrepancy may arise due to the higher H intake by Pd$_{GB}$.

After confirming the reversible hydriding and dehydriding behavior of Pd$_{GB}$ within the TEM chamber, we performed a high-resolution analysis of a region containing characteristic Σ3(111) GBs. **Figure 3c** shows a representative HR-TEM image overlaid with a y-direction strain map of Pd$_{GB}$ under 1 mbar N$_2$. In this region, three parallel Σ3(111) GBs (GB$_1$, GB$_2$, and GB$_3$) are identified. GB$_1$ and GB$_2$ are separated by 15 atomic layers, whereas GB$_2$ and GB$_3$ are closer, with only 4 layers between them. The strain map was generated via geometric phase analysis,[57] using average experimental lattice parameters of 2.398 Å for Pd(111) and 2.087 Å for Pd(002), both measured under non-H$_2$ conditions at the same HR-TEM magnification. The local strain within the region ranges from approximately −7% (compressive) to +3% (tensile), with an average strain of about −1% in this TEM image. Notably, strain tends to localize at the GBs, often as local maxima in either compression or tension. This is clearly illustrated in **Figure 3d**, which shows the averaged strain profile across a row of 10 Pd atoms as a function of vertical distance. The localized strain at GBs likely originates from the external forces imposed by neighboring NPs within the Pd network accumulating at the network's "connection points" (i.e., GBs).

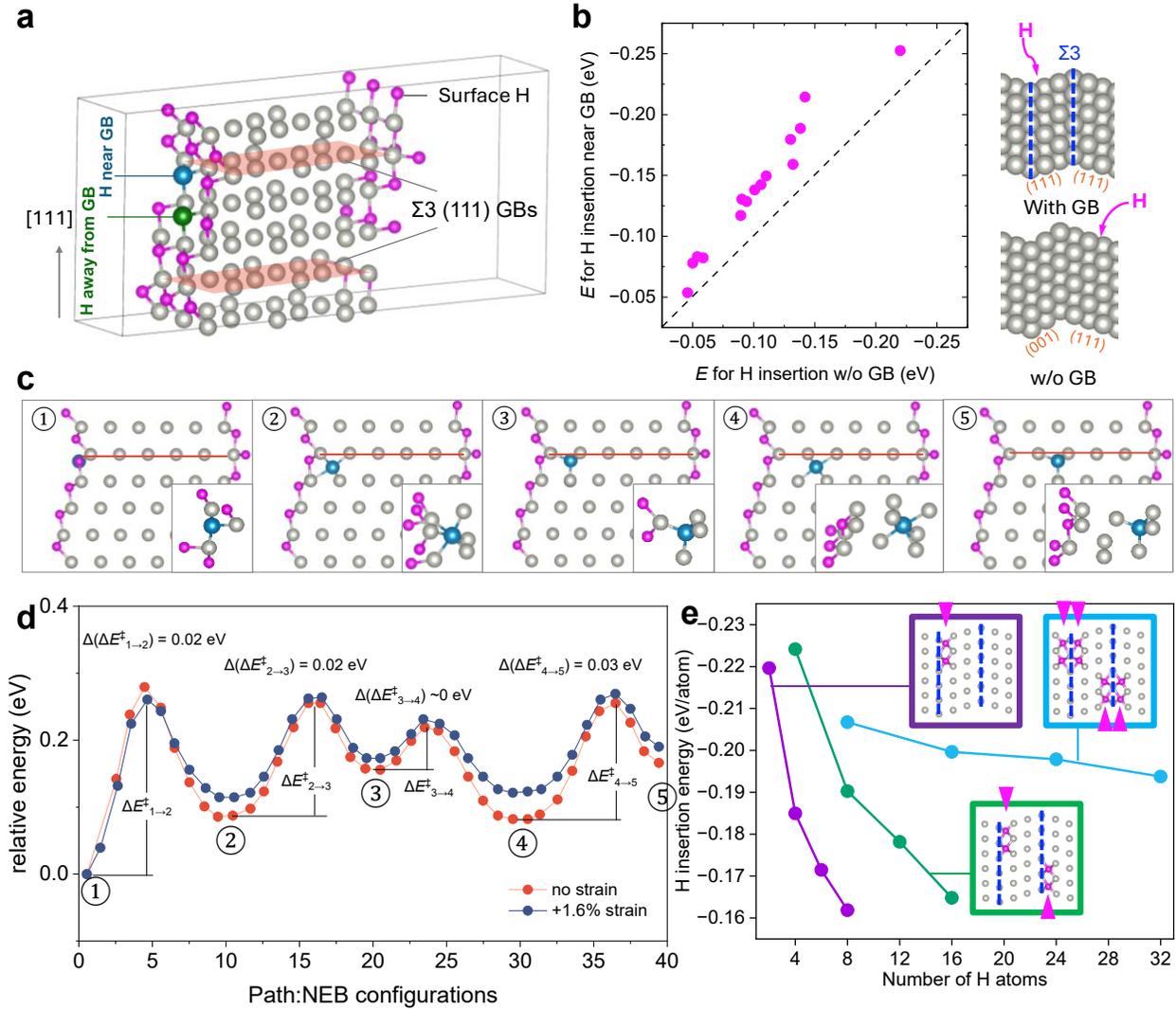

**Figure 4. Mechanism of H insertion into Pd$_{GB}$.** (a) Structural model of the GB-rich nanostructured Pd used in simulations of H insertion. Light peach plane: Σ3(111) GB; magenta: H; silver: Pd; blue: a H near a Σ3(111) GB; green: a H away from the GB. (b) Calculated energy difference for H insertion near the GB and without a GB. (c) Local atomic structures of interstitial H corresponding to the local energy minima on the calculated potential energy surface for hydrogen insertion along a Σ3(111) GB. Insets show the local coordination environment of the inserted H in each structure. (d) Potential energy profiles as a function of NEB configuration along the H insertion path, with and without 1.6% tensile strain along the [111] lattice direction. In both cases, the energy of the H adsorbed on the Pd surface is set to zero. (e) Average H insertion energy depending on the H content and insertion mode: along one plane near one GB (purple), along one plane near each GB (green), and along two planes near each GB (cyan).

Upon exposure to 1 mbar H$_2$, the Pd lattice undergoes expansion due to the hydrogen-induced phase transition. We attempted to capture this transition in real-time at the same location; however, the rapid nature of the phase change, combined with image drift at high magnification, made it challenging. As a result, we had to allow the system to equilibrate for approximately 20–30 minutes before acquiring new HR-TEM images. Consequently, instead of imaging the exact same region, we selected a nearby area containing similarly aligned parallel Σ3(111) GBs (two in close proximity to each other and the third farther away), comparable to those shown in **Figure 3c**, for structural comparison. **Figure 3d** shows its HR-TEM image overlaid with the corresponding strain map of Pd$_{GB}$ under 1 mbar H$_2$. The local strain ranges from approximately 0% (unstrained) to +4% (tensile), with an average strain of about +1%. This average strain increase upon H$_2$ exposure in this TEM image is consistent with the average lattice expansion of ~1.4% observed from the FFT diffraction pattern measurements in **Figures 3a** and **3b**. Most notably, the highest tensile strains of approximately +2.5% to +3% are localized at the GBs. The strain decreases rapidly with distance from the GBs. For instance, at point A in **Figure 3f**, just two atomic layers away from GB$_5$, the strain drops sharply from ~+2.5% at GB$_5$ to +0.3%. These findings clearly indicate that hydrogen preferentially inserts into Pd along Σ3(111) GBs rather than at non-GB regions.

The atomically well-defined GB structures in Pd$_{GB}$ enable us to construct a computational model to provide mechanistic insight into the experimentally observed behaviors. In this model, two parallel Σ3(111) GBs are separated by three atomic layers, and the exposed Pd surfaces are covered with a monolayer of adsorbed hydrogen atoms (**Figure 4a**; see **SI** for details). We compared the energy required for hydrogen insertion from a surface site near a GB (blue sphere) versus one farther away (green sphere). The results show that insertion from the site farther from the GB into the first Pd sublayer is ~0.1 eV less favorable than insertion adjacent to the GB (**Figure S10**). We also built a GB-free model with a similar nanostructure to the GB-rich model (**Figure 4b**). We calculated their H insertion energies for a range of configurations for comparison (Figures **4b**, **S11**, and **S12**). The result in **Figure 4b** shows that the H insertion near a GB is consistently more favorable than that at a similar location but in the GB-free region, supporting the conclusion that GBs promote H intake obtained from the experimental results.

The most favorable insertion pathway follows the GB plane, with the H atom alternating between tetrahedral and octahedral sites along the path (① to ⑤ in **Figure 4c**). The highest energy barrier along the insertion pathway is ~0.28 eV, corresponding to the first diffusion step from the surface site ① into the octahedral site ② in the first subsurface layer (**Figure 4d**). Transitions from octahedral to tetrahedral sites consistently show ~0.1 eV higher barriers compared to the reverse direction (for example, $\Delta E^{\ddagger}_{2\rightarrow 3} = 0.17$ eV vs $\Delta E^{\ddagger}_{3\rightarrow 4} = 0.06$ eV), which is consistent with the higher stability of H at the octahedral sites. We further found that all H configurations along the diffusion path are stabilized under tensile strain and destabilized under compressive strain (**Figure S14**). While an expanded Pd lattice stabilizes both octahedral and tetrahedral interstitial hydrogen atoms, the rate of this effect is larger for the tetrahedral sites. Correspondingly, the energy barriers for hydrogen transfer ($\Delta(\Delta E^{\ddagger})$) are reduced by ~0.02 to 0.03 eV under +1.6% tensile strain (**Figure 4d**).

Finally, we analyzed the energetics corresponding to three modalities of H insertion along the GBs (**Figure 4e**). Our calculations show that at the initial stages of H insertion, the energy gain is high in all three cases. However, as the amount of incorporated H increases, the energy gain due to H insertion along one plane near the GB decreases sharply, while the energy gain due to H insertion along both planes near the GBs remains nearly constant. This insertion process is accompanied by an increase in the Pd-Pd interplane spacing (**Figure S15**). Notably, H insertion increases this d-spacing on both sides of the mirror planes but does not affect the lattice beyond the immediate vicinity of the GB. This localized effect of H insertion, in particular the symmetric strain profile shown in **Figure S15c**, is consistent with the experimentally observed strain accumulation near the GBs (**Figure 3d**). Moreover, the corresponding H insertion mode, whereby H intake occurs on both sides of each GB (cyan in Figure 4e), is most thermodynamically preferred and is consistent with the observed rapid intake of H in GB-rich Pd.

In conclusion, we examined the response of nanostructured Pd with well-defined Σ3(111) GBs (i.e., Pd$_{GB}$) to hydrogen environment using in situ XRD and environmental TEM. Compared to Pd NPs with similar crystallite sizes (both with and without ligands), Pd$_{GB}$ exhibited significantly faster hydriding and dehydriding kinetics, indicating that GBs facilitate hydride formation. In situ TEM provided microscopic insight into the role of GBs, revealing localized tensile and compressive strains at GB sites even in the absence of H$_2$. Upon H$_2$ exposure, the tensile strain increased and remained highly localized, decaying within just a couple of atomic layers from the GBs. These observations suggest that GBs serve as preferential sites for hydrogen insertion.

Computational modeling using a structurally analogous system provides further insight into the mechanisms of H insertion, effects of external strain, and the character of H-induced strain distribution. The simulations showed that hydrogen insertion near the GB is energetically more favorable than at sites away from it, and that tensile strain at the GBs lowers the insertion barriers by ~0.02–0.03 eV. In addition, H insertion energetically prefers propagating along both planes near a GB. Together, these findings provide new atomic-level insights into the role of GBs in Pd hydride formation, highlight the potential of leveraging GBs to design advanced Pd-based functional materials, and offer a path for the design of high-rate H-transfer materials.

## ASSOCIATED CONTENT

**Supporting Information**. Chemicals and materials, general experimental methods (including the synthesis of Pd$_{NP}$, Pd$_{NP/UV}$, and Pd$_{GB}$ and characterization methods), GB identification method, crystallite measurements, the ligand effect, and computational modeling details. This material is available free of charge via the Internet at http://pubs.acs.org.

## AUTHOR INFORMATION

**Corresponding Author**

\* **Dongsheng Li** - *Physical & Computational Sciences Directorate, Pacific Northwest National Laboratory, Richland, Washington 99352, United States*; https://orcid.org/0000-0002-1030-146X; Email: Dongsheng.li2@pnnl.gov

**Peter V. Sushko** - *Physical & Computational Sciences Directorate, Pacific Northwest National Laboratory, Richland, Washington 99354, United States*; https://orcid.org/0000-0001-7338-4146; *Email:* peter.sushko@pnnl.gov

**Long Luo** - *Department of Chemistry, University of Utah, 315 S 1400 E, Salt Lake City, Utah 84112, United States*; https://orcid.org/0000-0001-5771-6892; Email: long.luo@utah.edu

**Author Contributions**

The manuscript was written through the contributions of all authors. All authors have given approval to the final version of the manuscript.

† These authors contributed equally.

ACKNOWLEDGMENT

This research was supported by the U.S. Department of Energy (DOE), Office of Science, Office of Basic Energy Sciences (BES): Material Sciences and Engineering Division, Synthesis and Processing Science Program, FWP 78705. L.L. also gratefully acknowledges support from the University of Utah and the Alfred P. Sloan Foundation (Grant # FH-2023-20829) for supporting the lab setup. TEM was conducted in the William R. Wiley Environmental Molecular Sciences Laboratory (EMSL), a national scientific user facility sponsored by the DOE Office of Biological and Environmental Research located at PNNL. The in situ XRD experiments were performed on the beamline 12-ID-D with beam time award (DOI:10.46936/APS-182977/60011131) from the Advanced Photon Source, a U.S.

Table of Contents artwork

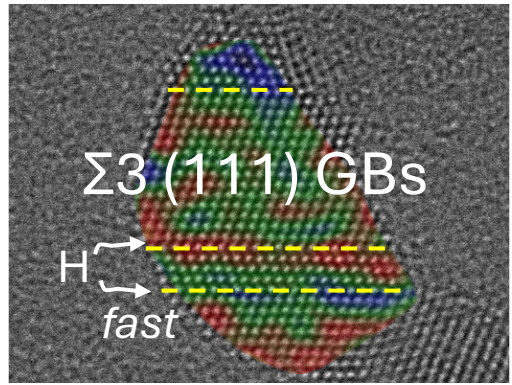

# Supporting Information

## ∑3 (111) Grain Boundaries Accelerate Hydrogen Insertion into Palladium Nanostructures

K. A. U. Madhushani,[1†] Hyoju Park,[2†] Hua Zhou,[3] Diptangshu Datta Mal,[1] Qin Pang,[2] Dongsheng Li,[2*] Peter V. Sushko,[2*] Long Luo[1*]

1. Department of Chemistry, University of Utah, Salt Lake City, Utah 84112, United States

2. Physical & Computational Sciences Directorate, Pacific Northwest National Laboratory, Richland, Washington 99354, United States

3. X-Ray Science Division, Argonne National Laboratory, Lemont, Illinois 60439, United States

Table of Contents

# Table of Figures

1. **Chemicals and Materials**

    Potassium tetrachloropalladate (II) (K$_2$PdCl$_4$, 98%), sodium citrate tribasic dihydrate (C$_6$H$_5$Na$_3$O$_7$·2H$_2$O, ≥ 99%), Acetone (C$_3$H$_6$O, ≥ 99.5%), and Pt foils were purchased from Sigma Aldrich. Sodium borohydride (NaBH$_4$, ≥ 98%) was acquired from MP Biomedicals. Sulfuric acid (H$_2$SO$_4$) and hydrochloric acid (HCl) were purchased from Fisher Scientific, and nitric acid (HNO$_3$) was purchased from VMR chemicals. An Ag/AgCl/saturated KCl reference electrode was acquired from CH Instruments. Deionized water (PURELAB, 18.2 MΩ cm, total organic carbon <3 ppb) was utilized throughout the experiments. All glassware used in the reactions was pre-cleaned with aqua regia (3:1 v/v mixture of HCl and HNO$_3$) prior to use.

2. **General Methods**

*2.1 Synthesis of Pd NPs (Pd$_{NP}$) and GB-rich assembly (Pd$_{GB}$).* Pd$_{NP}$ was synthesized as follows. First, 20.0 mL of 7.0 mM K$_2$PdCl$_4$ aqueous solution was added to 400.0 mL deionized water under constant stirring at room temperature. After 1 minute, 10.0 mL of 70.0 mM sodium citrate aqueous solution was added to the mixture. After another minute, 5.0 mL of a 60.0 mM aqueous solution of NaBH$_4$ was added as well. Upon the addition of NaBH$_4$, the solution immediately turned brown, indicating the formation of Pd NPs. The reaction mixture was stirred continuously for an additional 30 minutes to complete the synthesis, yielding the final Pd NP solution. Pd$_{GB}$ was prepared from Pd$_{NP}$ solutions using a previously reported method.[1] First, the concentration of the Pd NPs in the solution has been increased by a factor of 10 using a rotary evaporator. The preconcentrated Pd NP solution was then subjected to electrochemical processing to form assemblies using a three-electrode setup, consisting of a Pt foil (~240 mm²) as the working electrode, a Pt foil (~385 mm²) as the counter electrode, and an Ag/AgCl/saturated KCl electrode as the reference. Before use, the Pt foils were electrochemically cleaned in 0.5 M H$_2$SO$_4$ by cycling the electrode potential between -0.1 and 1.1 V vs RHE at a scan rate of 100 mV/s for 500 cycles using a potentiostat (CHI 650E, CH Instruments). For each synthesis, 8.0 mL of the concentrated Pd NP solution was used. An electrode potential of -2.0 V vs Ag/AgCl was applied to drive the assembly of Pd NPs. During this process, the NP solution gradually turned colorless over the course of several hours, and the process continued until the solution became completely colorless. The NP assembly was observed either as a precipitate at the bottom of the vial or adhered to the surface of the working electrode. The synthesized NP assembly was washed by adding and removing deionized water three times per day over a period of five days to completely remove the excess sodium citrate ligand. Following the cleaning process, XPS was performed to characterize the surface composition. The disappearance of the C 1s peak at 289 eV, corresponding to the C=O functional group of sodium citrate ligands, indicates that the ligands were effectively removed from the Pd NPs during the assembly and washing steps.

*2.2 Synthesis of dry Pd$_{GB}$ powder.* Dry Pd$_{GB}$ powder was used for XRD characterization. These powders were prepared from the wet Pd$_{GB}$ using the critical point drying (CPD) method. Before CPD, the solvent in the wet Pd$_{GB}$ was exchanged from DI water to acetone by carefully removing the supernatant. This solvent exchange step was repeated three times per day for 5 days. The wet Pd$_{GB}$ in acetone was then supercritically dried using an automated critical point dryer (EM CPD300, Leica). Acetone was first completely replaced with liquid CO$_2$ at 18 °C, and the liquid CO$_2$ was then evaporated by increasing the temperature to 37 °C. After fifty minutes, the CPD process was completed, resulting in dry Pd$_{GB}$ powder.

*2.3 Preparation of ligand-free Pd NPs (Pd$_{NP/UV}$).* Pd$_{NP/UV}$ was prepared by treating Pd$_{NP}$ with citrate ligands using a UV-ozone cleaner (Model L2002A3-US, Ossila Ltd). In our experiments, ~20 µL of concentrated Pd$_{NP}$ solution was drop-cast onto a microscope glass coverslip (Chase

Scientific, ZA0292) and subsequently treated in a UV ozone cleaner. Based on time-dependent studies, a 30-minute exposure was determined to be sufficient for complete removal of surface ligands from $Pd_{NP}$. Complete removal of citrate ligands from the Pd NP surface was achieved without altering the nanocrystal structure or composition, as confirmed by detailed analyses using HR-STEM (Figure S6), XPS (Figure S7), and XRD (Figure S8).

*2.4 Characterization*

a. *Ex-situ* TEM/STEM: Ex-situ TEM and STEM images were collected on a Talos F200X G2 S/TEM instrument at Lumigen Instrument Center, Wayne State University, and Thermo Fisher Titan 80-300 (image Cs-corrected) at Pacific Northwest National Laboratory. The particle size distribution was analyzed using ImageJ software (details in Figure S6). TEM samples were prepared by drop-casting ~10μL of each sample onto ultrathin C on easy carbon TEM grids. (Ted Pella Inc.).

b. *In-situ* TEM: In-situ gas flow TEM was conducted in an environmental TEM (Thermo Fisher Inc., Titan 80-300, image Cs-corrected) at 300 kV using a Gatan Metro 300 in-situ camera. $Pd_{GB}$ dispersed on a SiN TEM grid was first examined under ~ 1 mbar $N_2$ gas flow. The environment was then switched to ~1 mbar $H_2$ gas (99.9999% purity) and stabilized for ~30 min before imaging. Lattice parameters of $Pd_{GB}$ were re-examined 30 min after stopping the $H_2$ gas flow. Fast Fourier Transform (FFT) patterns were recorded from the $Pd_{GB}$ before, during, and after $H_2$ flow. Radial distributions of diffraction spots from the center spot were analyzed using ImageJ to show the change in lattice spacing across these three states. Strain maps were obtained from atomically resolved TEM images using geometric phase analysis (GPA) via Koch's FRWR tools plugin in Gatan DigitalMicrograph.[2] For the strain map under $H_2$ flow, reference lattice parameters of 2.398 Å ([111]) and 2.087 Å ([002])—average experimental values from the no-$H_2$ condition—were used to visualize lattice expansion under $H_2$ atmosphere.

c. *Ex-situ* XRD. XRD patterns were collected using a Bruker D8 Discover diffractometer housed in the University of Utah Nanofab Core Facility. The instrument was operated at 1600 W using a Cu Kα radiation source (λ = 1.5418 Å). Grazing Incidence X-ray Diffraction (GIXRD) was employed to enhance surface sensitivity and minimize signal interference from the silicon substrate. For analysis of $Pd_{NP}$ samples, ~50 μL of concentrated $Pd_{NP}$ solution was drop-cast onto a clean silicon wafer (Electron Microscopy Sciences, cat. no.71893-06) and allowed to dry under ambient conditions. For the $Pd_{NP/UV}$ sample, ~50 μL of the concentrated $Pd_{NP}$ solution was drop-cast onto a clean silicon wafer. Then, this sample was subjected to UV-ozone treatment. For $Pd_{GB}$ sample analysis, a thin layer of dry $Pd_{GB}$ powder is applied to the silicon wafer. The resulting samples were mounted on silicon-based zero-background holders for measurement.

d. *In-situ* XRD. High-resolution XRD measurements discussed in Figure 2 of the main text were carried out at the 12-ID-D beamline at the Advanced Photon Source of Argonne National Laboratory. The X-ray energy is 20 keV (wavelength: 0.61992 Å). X-ray geometry was calibrated using a standard $CeO_2$ reference powder to make the proper conversion between X-ray detector images and XRD patterns as a function of scattering angle 2*θ*. The XRD data in *Figure S8* were collected at NSLS-II 4ID ISR beamline at the Brookhaven National Laboratory. The X-ray energy is 17.7 keV (wavelength: 0.700452 A). The X-ray beam is focused by the beamline primary mirrors to provide a spot size down to approximately 40 μm (vertical) × 150 μm (horizontal).

e. XPS. XPS analysis was performed using a Thermo Fisher Scientific NEXSA UV and X-ray Photoelectron Spectrometer at the Lumigen Instrument Center, Wayne State University. Instrument control, data acquisition, and spectral analysis were conducted using Thermo Scientific Avantage software. All binding energies were calibrated to the C 1s peak at 284.8 eV. For the $Pd_{NP}$ sample, 20 µL of concentrated $Pd_{NP}$ solution was drop-cast onto a carbon paper substrate. The $Pd_{GB}$ sample was prepared by sonicating the wet $Pd_{GB}$ material in water, followed by drop-casting 20 µL of the resulting suspension onto carbon paper. For the $Pd_{NP/UV}$ sample, the particles were first resuspended from the glass coverslip (after UV-Ozone treatment) in ~100 µL of water, and 20 µL of this solution was then transferred onto carbon paper. All samples were dried overnight under vacuum prior to XPS analysis.

## 3. Σ3 (111) GB identification

The Σ3 (111) GB refers to a special type of coincidence site lattice boundary, where the two grains are related by a 60° rotation along <111> plane. This is also known as coherent twin boundary with the lowest energy grain boundary in fcc crystal structures. To identify Σ3(111) GB via high-resolution TEM, the crystal lattice image should be taken along <110> zone axis where (111) plane is visible. The Σ3 (111) GB can be identified where the atomic stacking sequence along across (111) direction is mirrored, like AB(C)BA, or where the orientation of rhombus shape of lattice is flipped (Figure S1). FFT analysis further confirms a 68° rotation between the grains across the Σ3 (111) GB.

In addition, Pd$_{GB}$ contains five-fold Σ3 GBs (Σ3$_1$ to Σ3$_5$) where five grains are arranged in a cyclic twin structure around a common axis along <110>. These twins often form a symmetrical pentagon shape (Figure S2), but also can be asymmetrical.[3] Five-fold Σ3 GBs gives additional geometrical internal strain due to angular misfit – theoretical angle of rhombus between (111) planes, 70.5 ° x 5 = 351.75 ° but adjusted to have a full circular angle 360°.

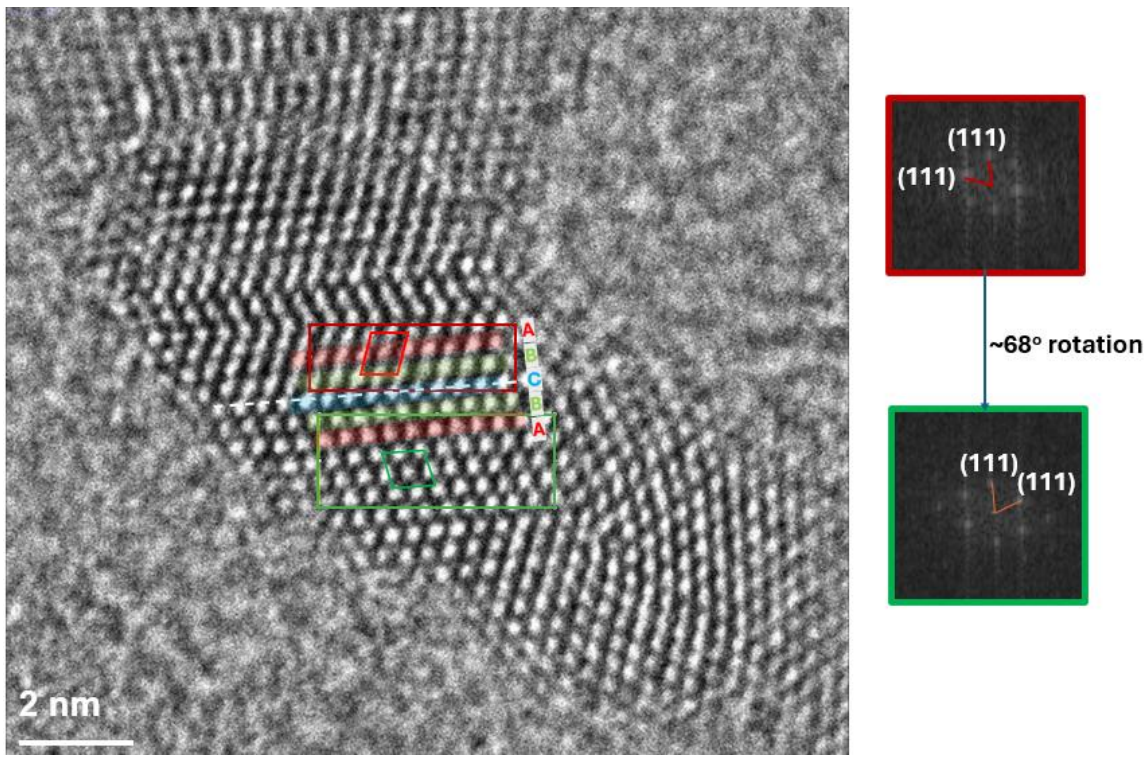

Figure S1. Identification of a Σ3 (111) Grain Boundary. The high-resolution TEM image was acquired along the <110> zone axis, revealing both (111) and (001) lattice planes. Fast Fourier transform (FFT) patterns from the two adjacent grains—highlighted by red and green boxes—are shown. The FFTs display an approximate 68° rotation between the two grains, a characteristic signature of a Σ3 (111) twin boundary.

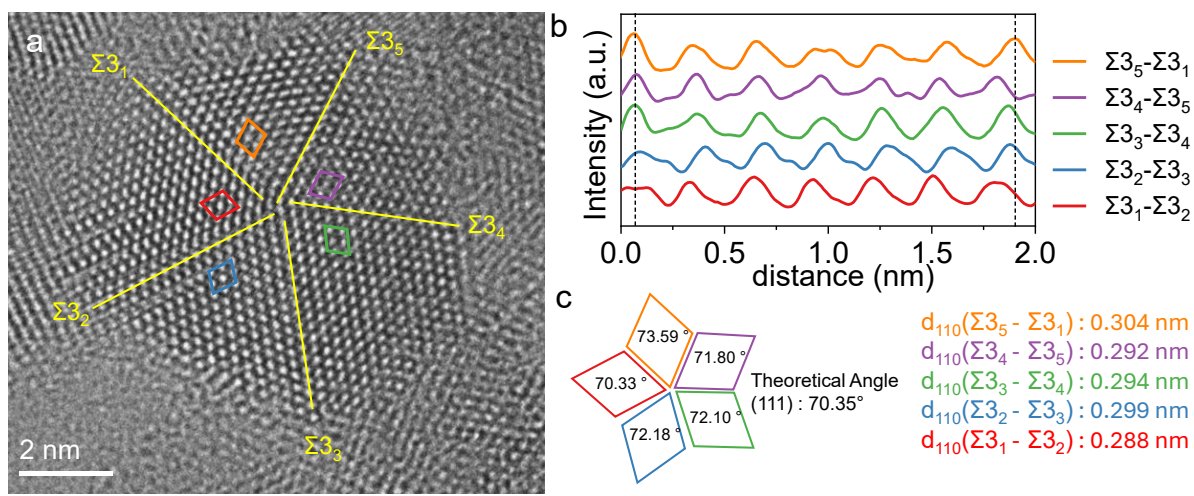

Figure S2. Assignment of a 5-fold twin comprising five symmetric Σ3 GBs (Σ3$_1$ to Σ3$_5$).

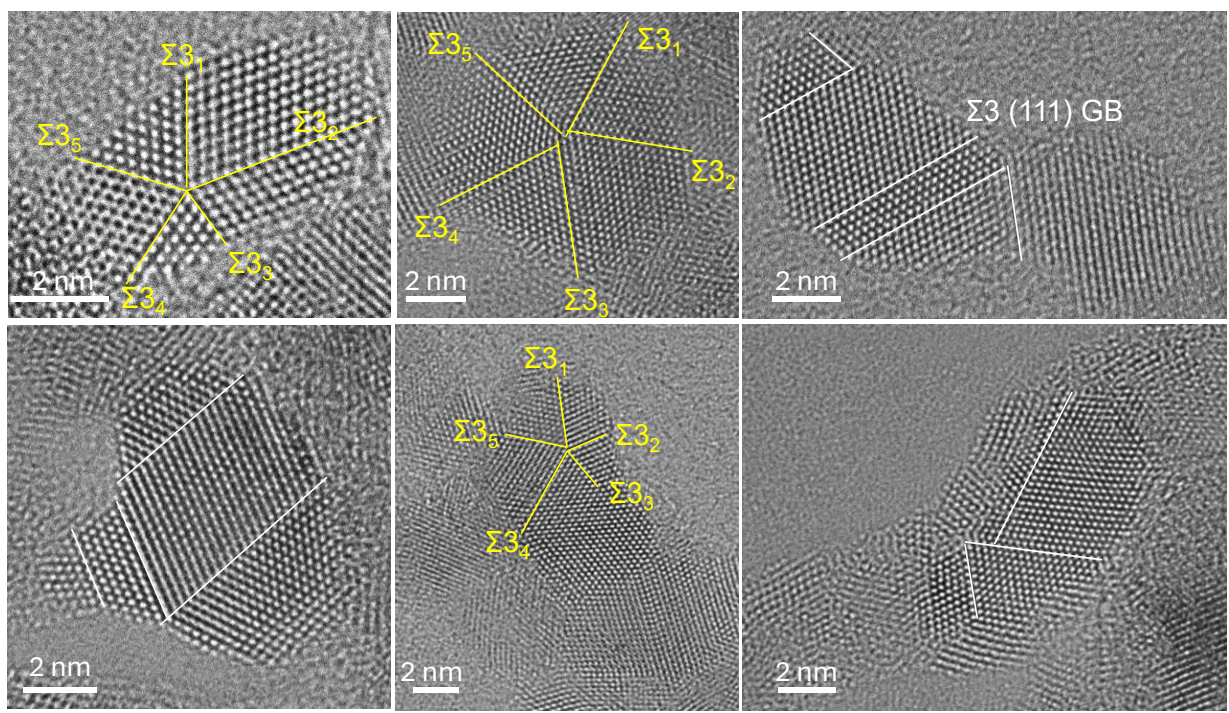

Figure S3. HR-TEM images of Pd$_{GB}$ showing abundant Σ3(111) GBs (highlighted by white lines) and five-fold twin structures with symmetric Σ3 GBs (highlighted by yellow lines).

## 4. Crystallite size measurements

The crystallite sizes of Pd$_{NP}$ and Pd$_{GB}$ were determined using both STEM and XRD techniques. For STEM analysis, particle diameters were directly measured from high-resolution STEM images using ImageJ software. A total of 100 nanoparticles or locations were randomly selected from different areas of the STEM image shown in Figure S4 to ensure a statistically meaningful size distribution.

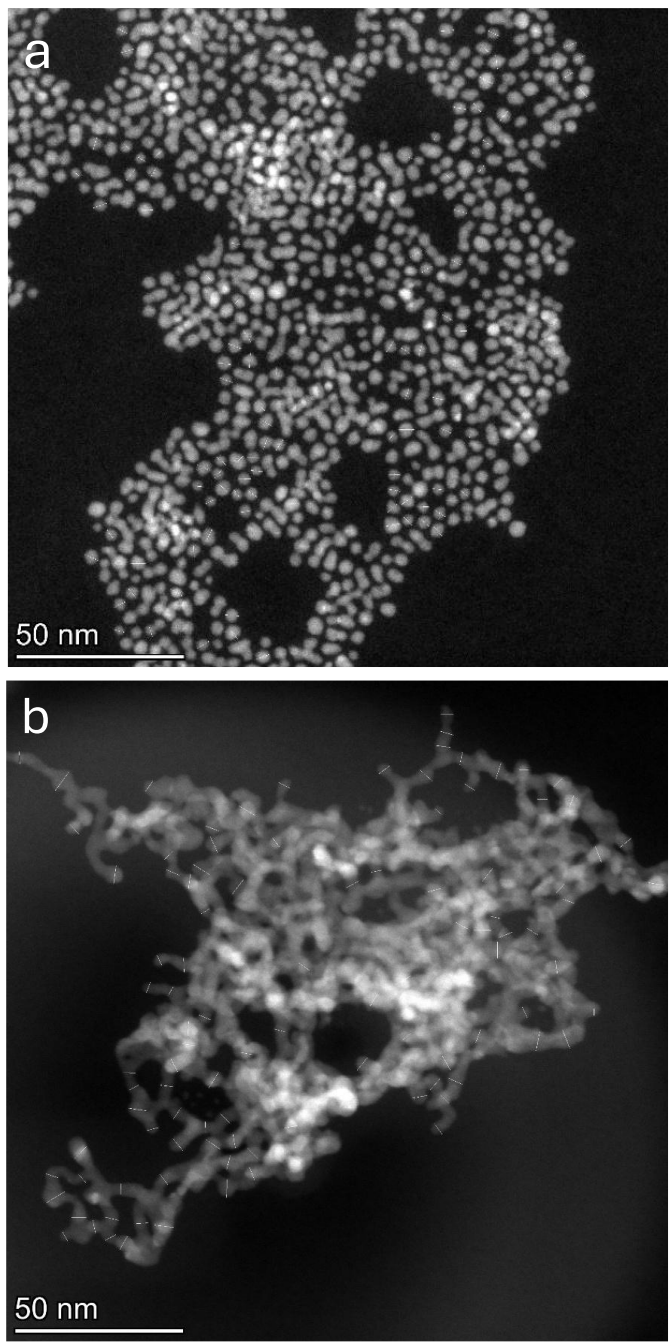

Figure S*4*. STEM images of (a) Pd$_{NP}$ and (b) Pd$_{GB}$ used for crystallite size analysis. The short straight lines indicate the sampling locations for size measurements.

In parallel, the crystallite sizes of $Pd_{NP}$ and $Pd_{GB}$ were estimated from the XRD peak width using the Scherrer equation[4, 5]

$$D = \frac{K\lambda}{\beta \cos\theta},$$

where $D$ is the crystallite size, $K$ is the shape factor (0.89), $\lambda$ is the X-ray wavelength (0.61992 Å), $\beta$ is the full width at half maximum (FWHM) of the diffraction peak in radians, and $\theta$ is the Bragg angle.

The crystallite sizes of $Pd_{NP}$ and $Pd_{GB}$ were estimated using the Pd (111) and (220) peaks from the high-resolution XRD pattern shown in Figure S5. Based on the (111) peak data (FWHM: 0.59° for $Pd_{NP}$ and 0.56° for $Pd_{GB}$), the calculated crystallite sizes are 5.4 nm and 5.7 nm, respectively. From the (220) peak data (FWHM: 0.90° for $Pd_{NP}$ and 0.85° for $Pd_{GB}$), the sizes are 3.6 nm and 3.8 nm, respectively. These XRD-derived sizes from (111) peak are consistently larger than those measured from STEM images and (220) XRD peak. The discrepancy arises from the fact that when two Pd NPs attach via their (111) facets, besides the formation of a Σ3(111) GB, an elongated structure can also form when the two NPs perfectly match, leading to broader crystalline domains detectable by XRD.

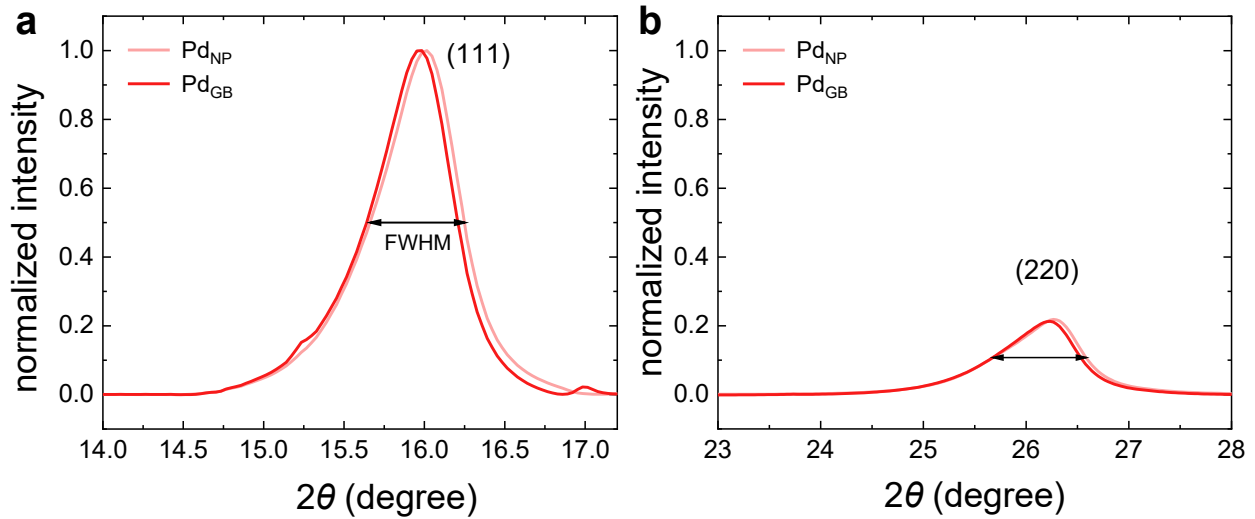

Figure S5. High-resolution XRD patterns showing (a) Pd(111) and (b) Pd(220) peaks for $Pd_{NP}$ and $Pd_{GB}$, which were used to estimate crystallite sizes using the Scherrer equation.

## 5. Structural characterization of ligand-free Pd NPs.

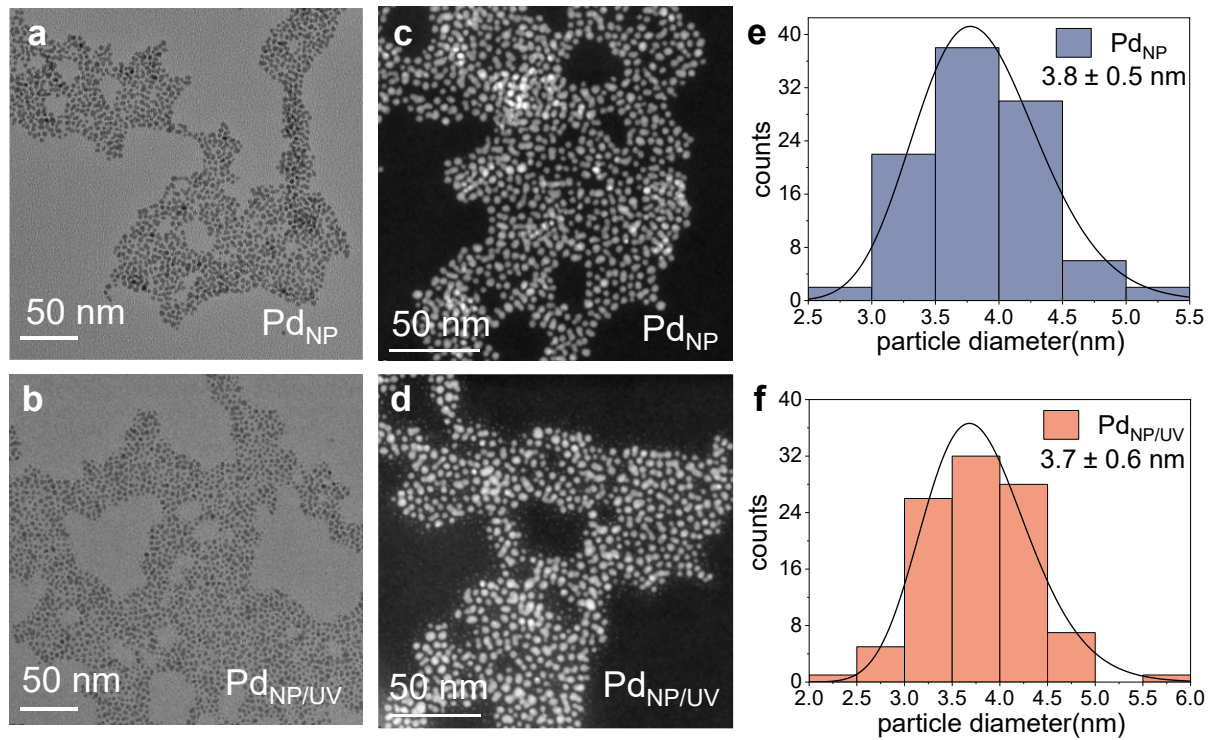

Figure S*6*. TEM and STEM images of (a) and (c) Pd$_{NP}$ with citrate ligands, (b) and (d) Pd$_{NP}$ after removing ligands using UV-Ozone cleaner (Pd$_{NP/UV}$), and (e) and (f) their corresponding particle diameter distribution curve obtained from analyzing 100 particles.

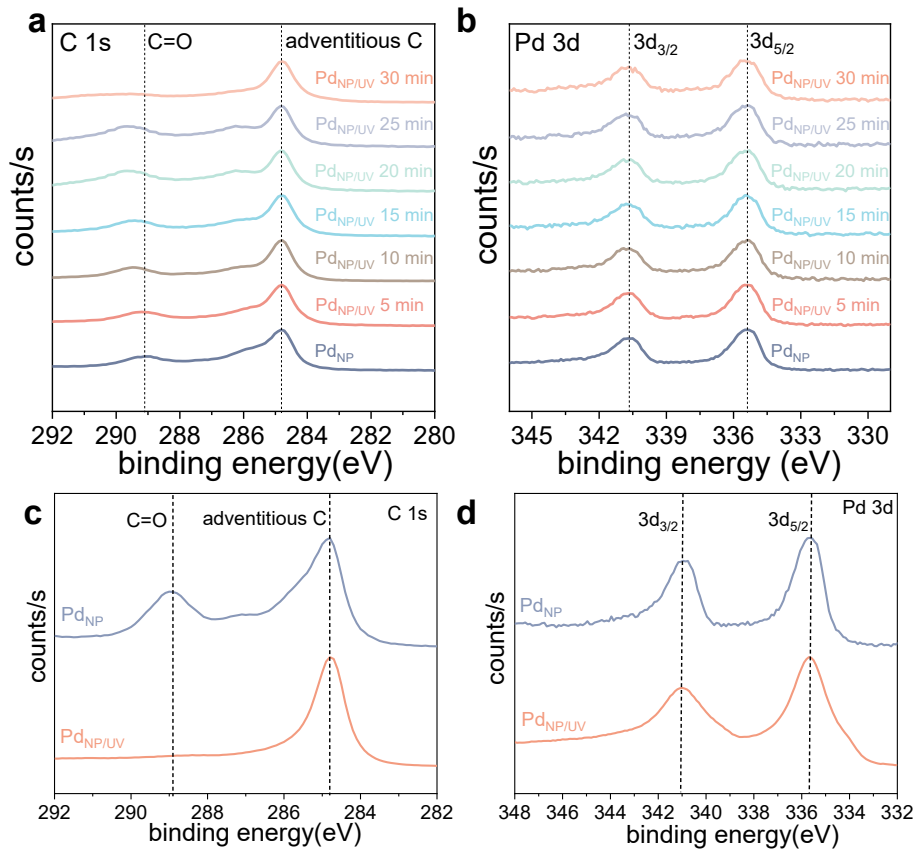

Figure S7. High-resolution XPS spectra of (a) C 1s and (b) Pd 3d for Pd$_{NP}$ subjected to UV-ozone treatment for 0 to 30 min. The gradual shift and eventual disappearance of the C 1s peak at ~288.5 eV indicates the removal of citrate ligands, which is complete after 30 min. (c) and (d) show direct comparisons of the C 1s and Pd 3d regions, respectively, for Pd$_{NP}$ before and after 30 min of UV-ozone treatment (Pd$_{NP/UV}$).

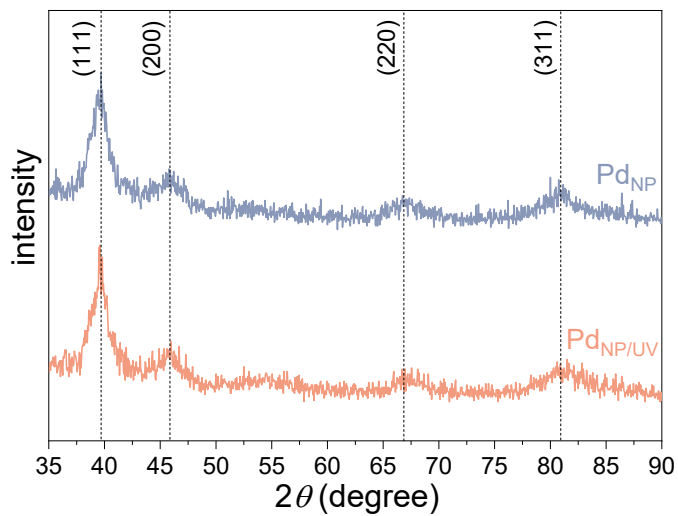

Figure S8. XRD patterns for Pd$_{NP}$ and Pd$_{NP/UV}$.

## 6. Effect of the ligands on the kinetics of H-induced transformation

To establish the effect of surface ligands on the kinetics of hydrogen incorporation, we analyzed the time-resolved evolution of the (111) Bragg peak intensity in samples with and without ligands using in-situ XRD. These experiments were performed on $Pd_{NP}$, $Pd_{NP/UV}$, and $Pd_{GB}$ under hydrogen gas flow and gas-switching conditions at the 4-ID ISR beamline of NSLS-II at Brookhaven National Laboratory. From this study, it was observed that the $Pd_{GB}$ sample exhibited faster hydrogen incorporation kinetics compared to both $Pd_{NP}$ and $Pd_{NP/UV}$ samples, further supporting our assertion that GBs play a dominant role in facilitating hydrogen uptake in Pd structures.

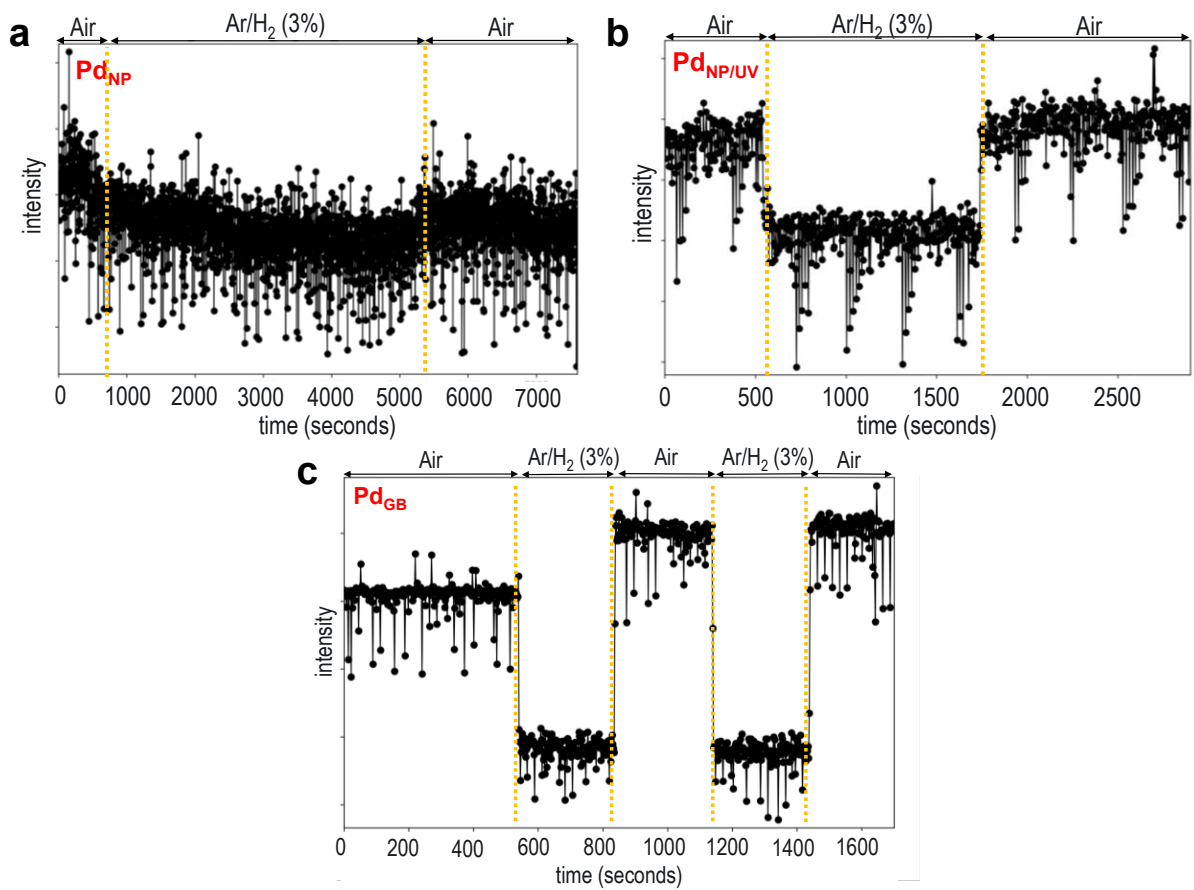

Figure S9. Integrated intensity of the Pd(111) Bragg peak as a function of time under alternating flows of air and 3% $H_2$ in Ar for (a) $Pd_{NP}$, (b) $Pd_{NP/UV}$, and (c) $Pd_{GB}$. The observed glitches are attributed to the beamline optics/mono instability and accelerator ring fill patterns, not from the sample itself, during the measurements.

## 7. Computational modeling

The energetics of H interactions with nanostructured Pd and the effect of H incorporation on the strain and charge density distribution were investigated using ab initio simulations within the periodic slab model. The GB-rich Pd was represented using a 72-atom periodic slab model containing two Σ3 (111) twin boundaries. The slab was terminated with Pd(111) planes, resulting in a corrugated surface, as shown in Figure 4 of the main text. For comparison, we also considered a slab model free of GBs. In this case, the slab was terminated by (111) and (100) planes (Figures S10 and S11). In both cases, the slabs were modelled using nearly orthorhombic supercells with the supercell parameters of approximately 33 Å, 5.5 Å, and 13.5 Å along the [11-2], [1-10], and [111] Pd lattice vectors, respectively.

The calculations were performed within the density functional theory (DFT) formalism as implemented in the VASP code.[6, 7] The Perdew-Burke-Ernzerhof (PBE) exchange-correlation functional[8] and projected augmented wave potentials were used throughout.[9] The plane wave basis set cutoff was 400 eV; the Γ-centered k-mesh 1×3×2 was used in most calculations. The total energy convergence criterion was set to $10^{-5}$ eV. The charge density distribution was analyzed using the Bader approach.[10]

To analyze the effect of Σ3 (111) GBs on the stability of interstitial H, we compared insertion energies for H located near the GBs and far from the GB (Figure S10). The GB-free model was also used for comparison (Figure S11 and Figure S*12*). Insertion energies of H atoms into a partially hydrogenated Pd slab (Pd$_{72}$) were calculated relative to the energy of the gas-phase H$_2$ molecule:

$$E_{\text{in}} = E[\text{Pd}_{72}\text{H}_{n+1} \text{ slab}] - (E[\text{Pd}_{72}\text{H}_n \text{ slab}] + \frac{1}{2}E[\text{H}_2])$$

The average H insertion energies are defined as

$$E_{\text{in-a}} = \frac{E[\text{Pd}_{72}\text{H}_n] - E[\text{Pd}_{72}] + \frac{n}{2}E[\text{H}_2]}{n}$$

According to this definition, $E_{\text{in-a}}$ values are negative if H insertion is thermodynamically preferable.

We considered a range of interstitial concentrations (up to 100% octahedral site occupancy per plane) and configurations. In all cases, the total energies of the systems were minimized with respect to the internal coordinates of all atoms and the supercell parameters under the fixed volume condition. Comparison of the average H insertion energies shows in Figure 4b that interstitial H is more stable near the Σ3 (111) GBs.

We examined the convergence of the calculated energies of the H interaction with Pd with respect to the k-mesh density and the plane wave cutoff. These calculations (see Figure S13) indicate that the k-mesh 1×3×2 results in an ~0.05 eV overestimation of the H insertion energies.

To construct the potential energy surface for H in-diffusion, we first determine the configurations and relative energies of local H minima in the octahedral and tetrahedral interstitial sites in the vicinity of the GB, as shown in **Figure 4e**. Then, diffusion paths and barriers for H transfer between these minima were calculated using the climbing-image nudged elastic band method (NEB) and 8 NEB images between each minimum (Figure 4d). In these calculations, we considered the regime of a high concentration of H pre-adsorbed on the surface; their presence induces an ~1% lateral expansion in our model system. The same calculations were performed for the strained Pd, where the uniaxial strain was applied by increasing the slab supercell parameter c corresponding to the [111] lattice vector.

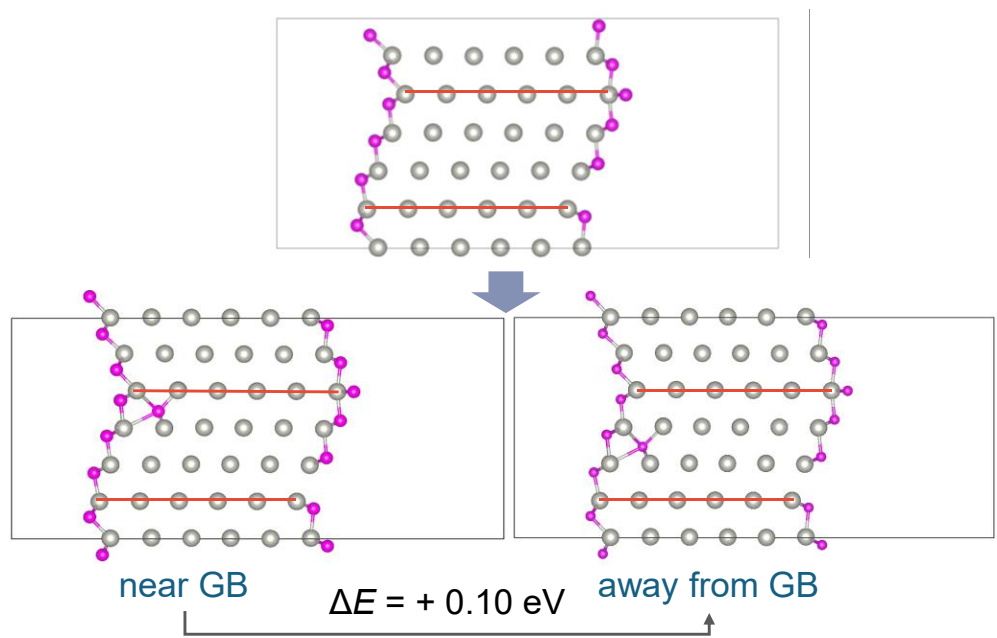

Figure S10. The calculated energy difference of 0.1 eV for H insertion near vs away from the GB suggests that H accumulation near the GB is more thermodynamically preferred than their dispersion over Pd. Here, the surface of the Pd slab was modified to account for the presence of H adsorbed on the surface. For the system used here, surface H coverage results in an ~1% lateral supercell expansion

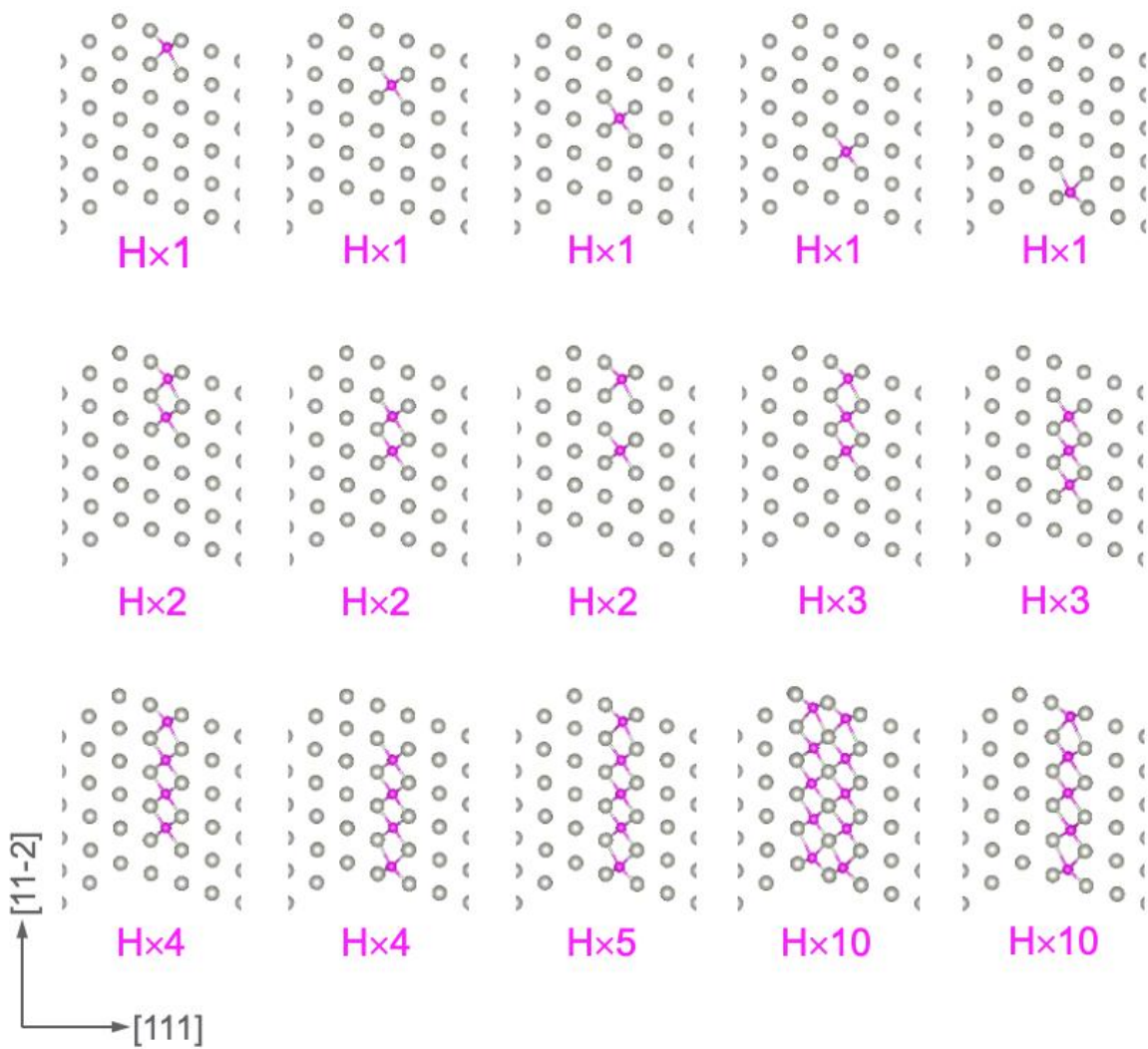

Figure S11. Configurations of interstitial H atoms (magenta) in GB-free $Pd_{72}$ (grey) slab model used to evaluate the stability of subsurface hydrogen and the magnitude of H-induced strain. The labels indicate the number of incorporated H atoms. The number of interstitial octahedral sites between two consecutive (111) planes is 10. Accordingly, the two configurations with the largest number of H atoms (H×10) correspond to 50% and 100% occupancy of the octahedral interstitial sites per plane.

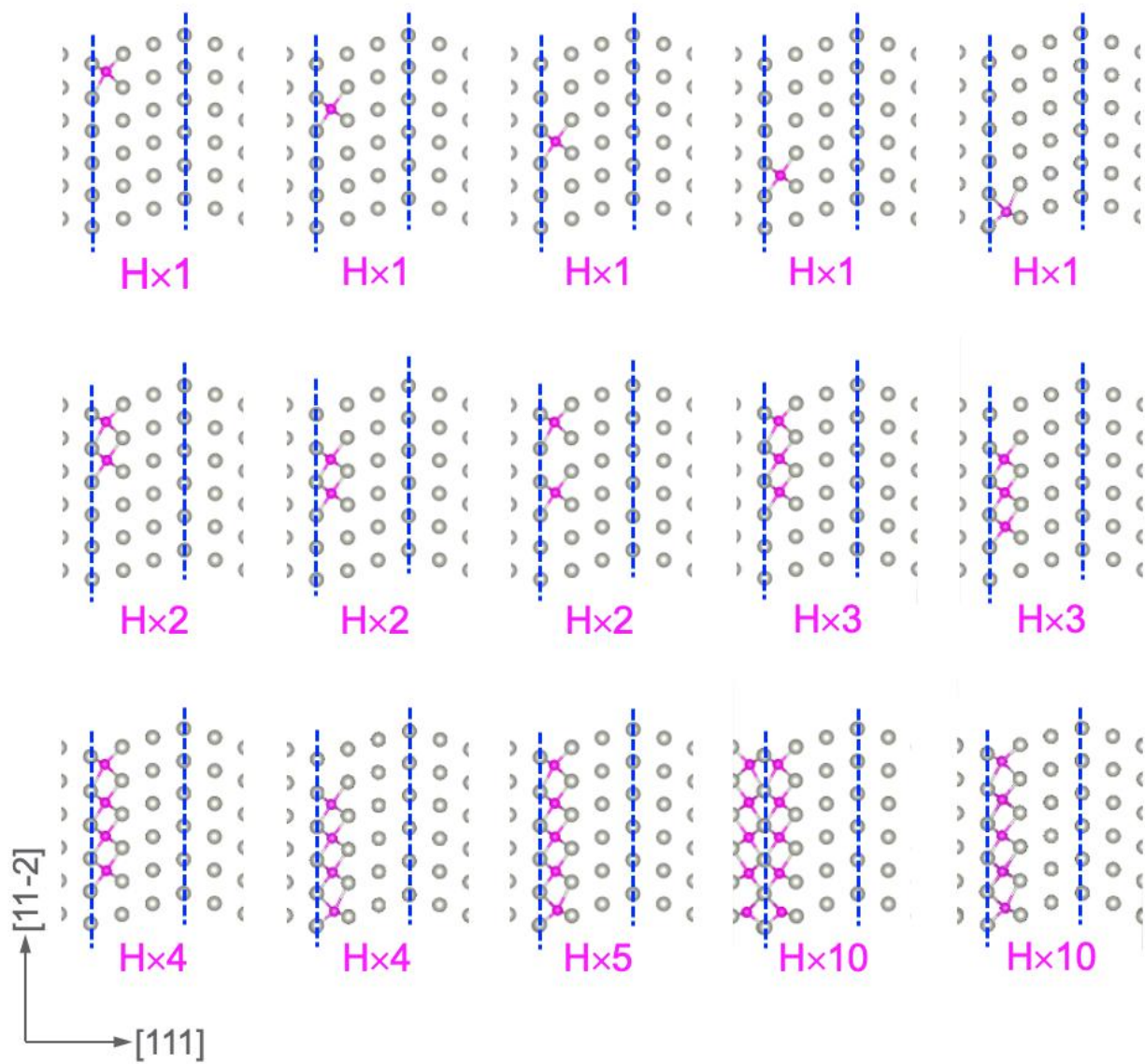

Figure S12. Configurations of interstitial H atoms (magenta) in GB-containing $Pd_{72}$ (grey) slab model used to evaluate the stability of subsurface hydrogen and the magnitude of H-induced strain. The locations of mirror planes associated with the $\Sigma3(111)$ grain boundaries are indicated with dashed blue lines. The labels indicate the number of incorporated H atoms. The number of interstitial octahedral sites between two consecutive (111) planes is 10. Accordingly, the two configurations with the largest number of H atoms (H×10) correspond to 50% and 100% occupancy of the octahedral interstitial sites per plane.

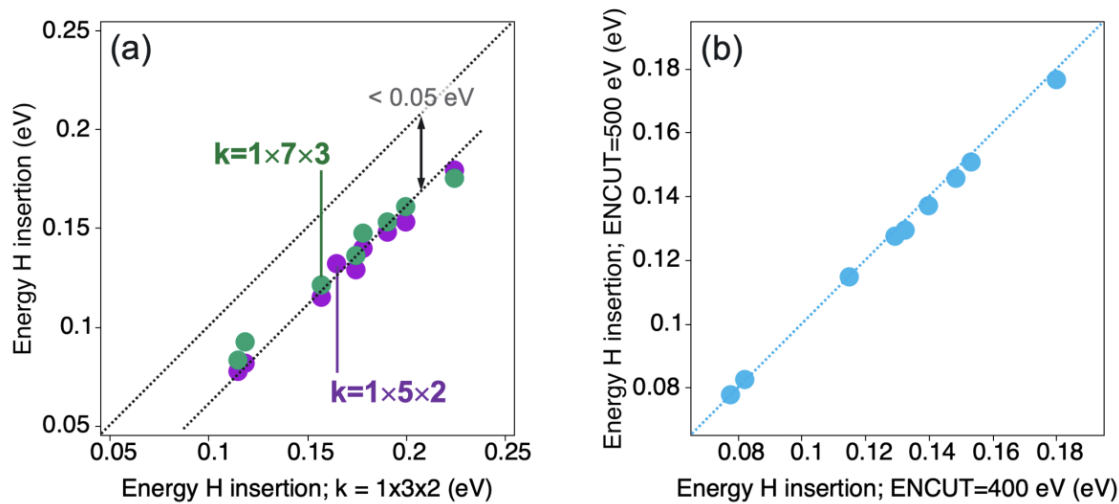

Figure S13. Calibration of the calculated average H insertion energies vs the density of the k-mesh (a) and plane-wave basis set cutoff energy ENCUT (b) for selected configurations of up to 16 H atoms in the $Pd_{72}$ slab model containing two $\Sigma3$ GBs. (a) The average insertion energies calculated for the 1×5×2 and 1×7×3 k-meshes are nearly identical and deviate by less than 0.05 eV from the energies calculated using the 1×3×2 k-mesh. The plane wave basis cutoff energy was 400 eV in all cases. (b) Correlation of the H insertion energies calculated using the 1×5×2 k-mesh and 400 eV and 500 eV plane-wave basis set cutoffs. The nearly identical values of the insertion energies indicate that the 400 eV cutoff provides sufficient accuracy for these calculations.

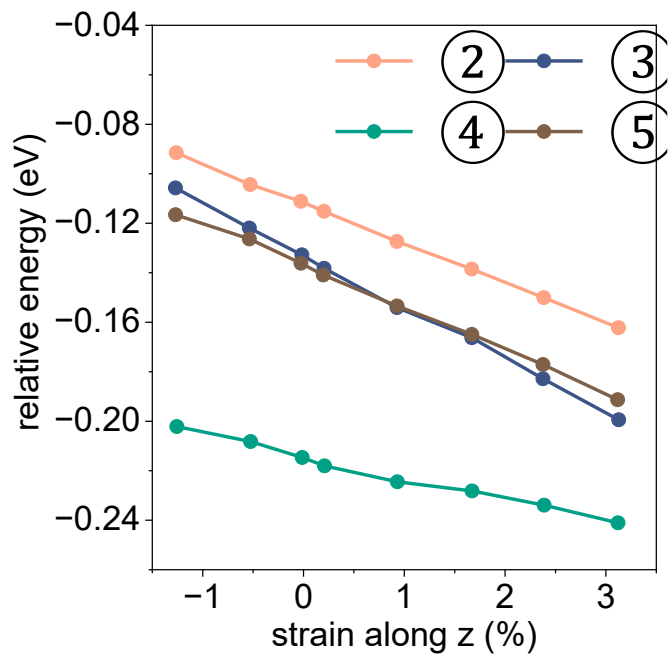

Figure S14. Relative energies for the structures ② to ⑤ shown in Figure 4c in the main text as a function of strain along the [111] lattice direction (z). The tensile strain stabilizes interstitial H in both octahedral and tetrahedral configurations.

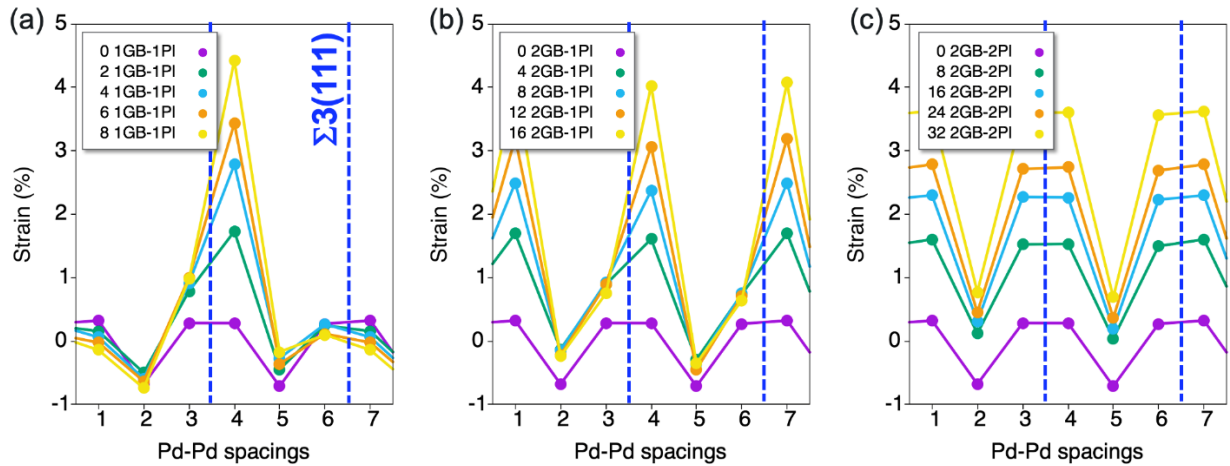

Figure S15. Simulated effect of interstitial H on the distances between Pd planes along the [111] lattice direction depending on the H insertion mode (Figure 4, main text): (a) H in a single plane near a GB; (b) H in a single plane near two GBs; (c) H in two planes near each GB. The strain magnitude was calculated relative to the average Pd-Pd distances in the H-free Pd slab. In the absence of interstitial H, the interplane Pd-Pd distances near the Σ3(111) mirror planes (indicated with the blue dashed lines) are ~1% larger than the Pd-Pd distance away from the mirror plane. The strain magnitudes of the H-containing regions are comparable, irrespective of the H insertion mode. The asymmetric profiles in (a) and (b) show that interstitial H affects the spacing between Pd planes near the mirror plane but not in the plane away from the mirror plane.